\documentclass[doublecol]{epl2}

\title{The resilience of interdependent transportation networks under targeted attack}
\shorttitle{The resilience of interdependent transportation networks under targeted attack} 

\author{Peng Zhang\inst{1} \and Baisong Cheng\inst{1} \and Zhuang Zhao\inst{1} \and Daqing Li\inst{2,3}\footnote{daqingl@buaa.edu.cn} \and Guangquan Lu\inst{4}\footnote{lugq@buaa.edu.cn} \and Yunpeng Wang\inst{4} \and Jinghua Xiao\inst{1}}
\shortauthor{Peng Zhang\etal}

\institute{
  \inst{1} School of Science, Beijing University of Posts and Telecommunications, Beijing 100876, China\\
  \inst{2} School of Reliability and Systems Engineering, Beihang University, Beijing 100191, China\\
  \inst{3} Science and Technology on Reliability and Environmental Engineering Laboratory, Beijing 100191, China\\
  \inst{4} Beijing Key Laboratory for Cooperative Vehicle Infrastructure Systems and Safety Control, Beihang University, Beijing 100191, China
}
\pacs{89.75.Fb}{Structures and organization in complex systems}
\pacs{05.10.-a}{Computational methods in statistical physics and nonlinear dynamics}
\pacs{89.40.-a}{Transportation}

\abstract{
Modern world builds on the resilience of interdependent infrastructures characterized as complex networks. Recently, a framework for analysis of interdependent networks has been developed to explain the mechanism of resilience in interdependent networks. Here we extend this interdependent network model by considering flows in the networks and study the system's resilience under different attack strategies. In our model, nodes may fail due to either overload or loss of interdependency. Under the interaction between these two failure mechanisms, it is shown that interdependent scale-free networks show extreme vulnerability. The resilience of interdependent SF networks is found in our simulation much smaller than single SF network or interdependent SF networks without flows.}
\begin{document}

\maketitle

\section{Introduction}

Over the past decade, network theory has become one of the major tools in studying complex systems, and has been proved useful for the description and analysis of complex systems in various fields \cite{1-1,1-2,1-3,1-4,1-5,1-6,1-7,1-8,1-9,1-10,1-11}. As one of the most fundamental questions, the resilience of networks has been studied intensively, which helps to provide efficient solutions to protect real networks against faults or attacks \cite{1,101,2,3,4}. However, most of these studies are based on the assumption that networks are isolated, neglecting the interdependency between them. Recently, Buldyrev $et\ al$. presented a theory of interdependent networks. In their work, a mutual percolation model is proposed to study the vulnerability of network of networks where the links between the networks are interdependency links \cite{5}. This has initiated a series of studies about various interdependent networks. Parshani $et\ al$. proposed a theoretical framework for studying the case of partially interdependent networks by defining the coupling strength $q$. Their findings showed that reducing the coupling strength leads to a change from first to second order percolation transition \cite{5-1}. Gao $et\ al$. has generalized the theory on network of networks, showing that the percolation theory of a single network is a limiting case of this generalization \cite{5-2,5-3,5-4}.

Interdependency does exist between various networks for transferring flows including airlines, urban road networks, Internet and power grid \cite{6,7,8,9,10,10-1,10-2}. The failure of network components, either by random breakdown or intentional attacks, could change the balance of flows causing overloads and trigger the cascading failures, which probably ends up with catastrophes \cite{1, 101,11, 12, 25, 26, 27}. In single networks, Motter $et\ al$. presented a model to consider a cascade of overload failures \cite{1,101}. In this paper, we generalize it to interdependent networks, study their resilience against cascading failures caused by both overloads and by loss of interdependency under random or intentional attack strategies. Firstly we study the resilience of interdependent scale-free networks under different attacks and explore in detail how the resilience depend on scaling exponent $\gamma$ of degree distributions. Secondly, we study how the interdependent Erd\H{o}s-R\'{e}nyi networks respond to different attack strategies. Finally, we perform a study to understand the effect of interdependency $q$ on the coupled networks resilience.

\section{The Model}
\label{Cascading Model}

In our model, we assume for simplicity that the coupled networks, $A$ and $B$, have the same size $N_{A}=N_{B}=N$ and same degree distribution, $P_{A}(k)=P_{B}(k)$. $q$ quantifies the fraction of nodes having interdependency link to networks $B$. Each node in network $A$ depends only on one node in network $B$ and vice versa, which establishes a one-to-one bidirectional dependent relation. The one-to-one bidirectional dependent links are established randomly to avoid any correlations among the two networks. In the iterative failure process, if node $A_{i}$ stops functioning due to attack or overload failure, node $B_{i}$ which depends on $A_{i}$ stops functioning as well, and vice versa. The load quantifies the amount of flows that a node is requested to transmit and is considered to depend on the total number of shortest paths passing through it \cite{13,14,15}. The load of node $i$ can be denoted by
\begin{equation}
 L(i)=\sum_{(v_{1},v_{2})}\frac{\sigma_{v_{1},v_{2}}(i)}{\sigma_{v_{1},v_{2}}} \ \ \ \ i=1,2,\cdots,N
\end{equation}
Where $\sigma_{v_{1},v_{2}}$ is the total number of shortest paths between node $v_{1}$ and $v_{2}$, $\sigma_{v_{1},v_{2}}(i)$ is the number of shortest paths between node $v_{1}$ and $v_{2}$ through node $i$.

Following Ref. \cite{1}, the capacity of node $i$ is denoted by
\begin{equation}
 C(i)=(1+\alpha)\ast L_{0}(i) \ \ \ \ i=1,2,\cdots,N
\end{equation}
Where $\alpha$ is the tolerance parameter, and $L_{0}(i)$ is the initial load of node $i$. A node is failed when its load exceeds the capacity.

If we remove intentionally ('attack') some nodes in network $A$, firstly it will induce cascading failure by the redistribution of loads among the nodes in network $A$. Here we assume that only the nodes in the giant component remain functional. The failed nodes may disintegrate network $A$ and all nodes outside the giant component will cause their dependency counterparts in network $B$ to fail. These failed nodes in network $B$ will cause overloads and more nodes to fail. This process will continue recursively until no further damage is produced either by overloads or by interdependency losses. In our model, the links between network $A$ and $B$ only reflect the dependence relationship. It is different from reference \cite{20}, where links between network $A$ and $B$ is used for traffic process.

Different attack strategies have been studied in isolated networks with or without considering overloads Refs.\cite{1,101,16,17,18,19,19-1,19-2}. We here focus on three different attack strategies: (i) remove the node with highest load; (ii) remove the node with largest degree; (iii) remove a node randomly. We compare here these three different strategies.

\section{Results }\label{ER-SF}

Now we present numerical simulations obtained for different attack strategies on interdependent ER networks (ER-ER) and interdependent scale-free networks (SF-SF) with flows. To generate SF-SF networks, we use the method mentioned in Ref. \cite{1-7}. By this method, we could compare SF network with different scaling exponent $\gamma$ by fixing the average degree $<k>$. The relative size of the largest connected component $G=N'/N$ is used to quantify the network resilience, where $N$ and $N'$ are the size of largest component before and after cascading respectively. Considering the computation cost from overload calculation, the network size in the simulation is $N_{A}=N_{B}=5000$. We intentionally remove one node with the highest load or the largest degree, which are compared with random removal. After the initial attack, node can fail in a domino-like process due to either overload inside one network or loss of interdependency between two networks. Therefore, the cascading failure process is more complicated in the interdependent networks when these two failure mechanisms are interacting and receiving different feedback.

\begin{figure}
\centering
\includegraphics[width=8cm]{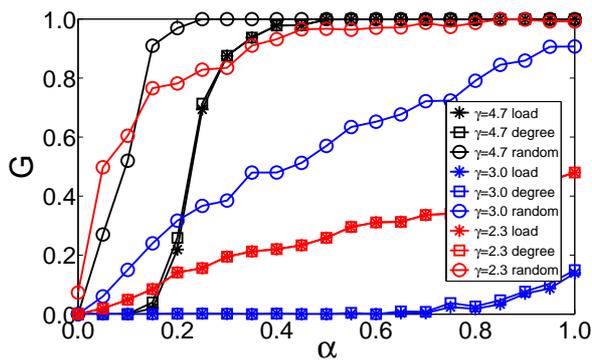}\\
\caption{The relative size of giant component $G$ of a system of on interdependent SF networks as a function of the tolerance parameters $\alpha$, when $N_{A}=N_{B}=5000$, $<k>\approx4$ and $q$=1 for three different scaling exponents $\gamma$ under different attack strategies. 
}\label{SF4}
\end{figure}

We begin the study with interdependent scale-free networks. It is shown in Fig.~\ref{SF4} that intentional attack makes more damage to interdependent networks than random removal. Interdependent scale-free network with $\gamma=3$ is found to be more vulnerable than other interdependent networks ($\gamma=2.3$ and $4.7$). Due to the correlation between degree and load \cite{14}, the attack based on nodes' degree is found in our study as harmful as attack based on nodes' load. In the following we will not specify the attack types (based on degree or load) when we refer to attack.

\begin{figure}
\centering
\includegraphics[scale=0.13]{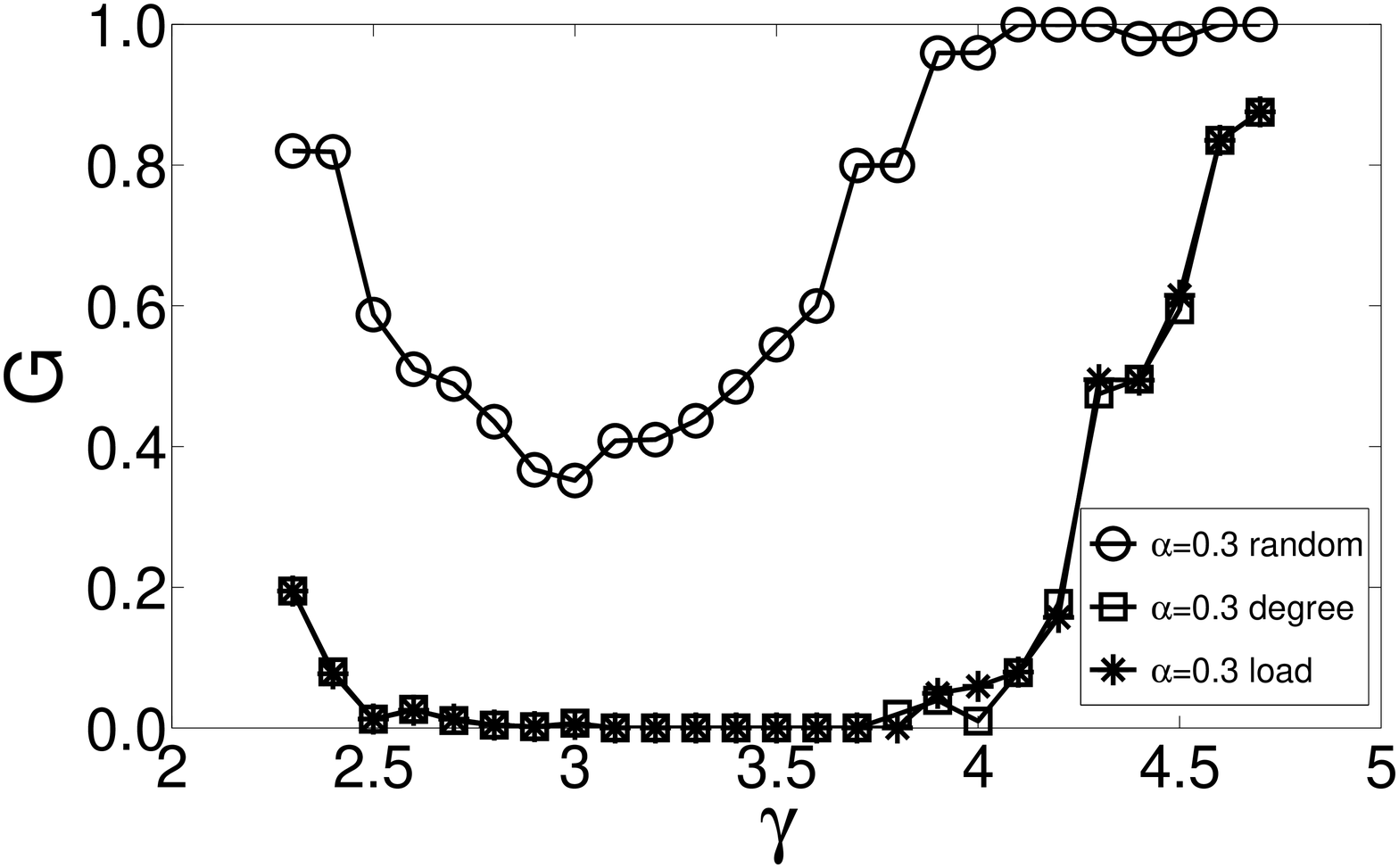}\includegraphics[scale=0.13]{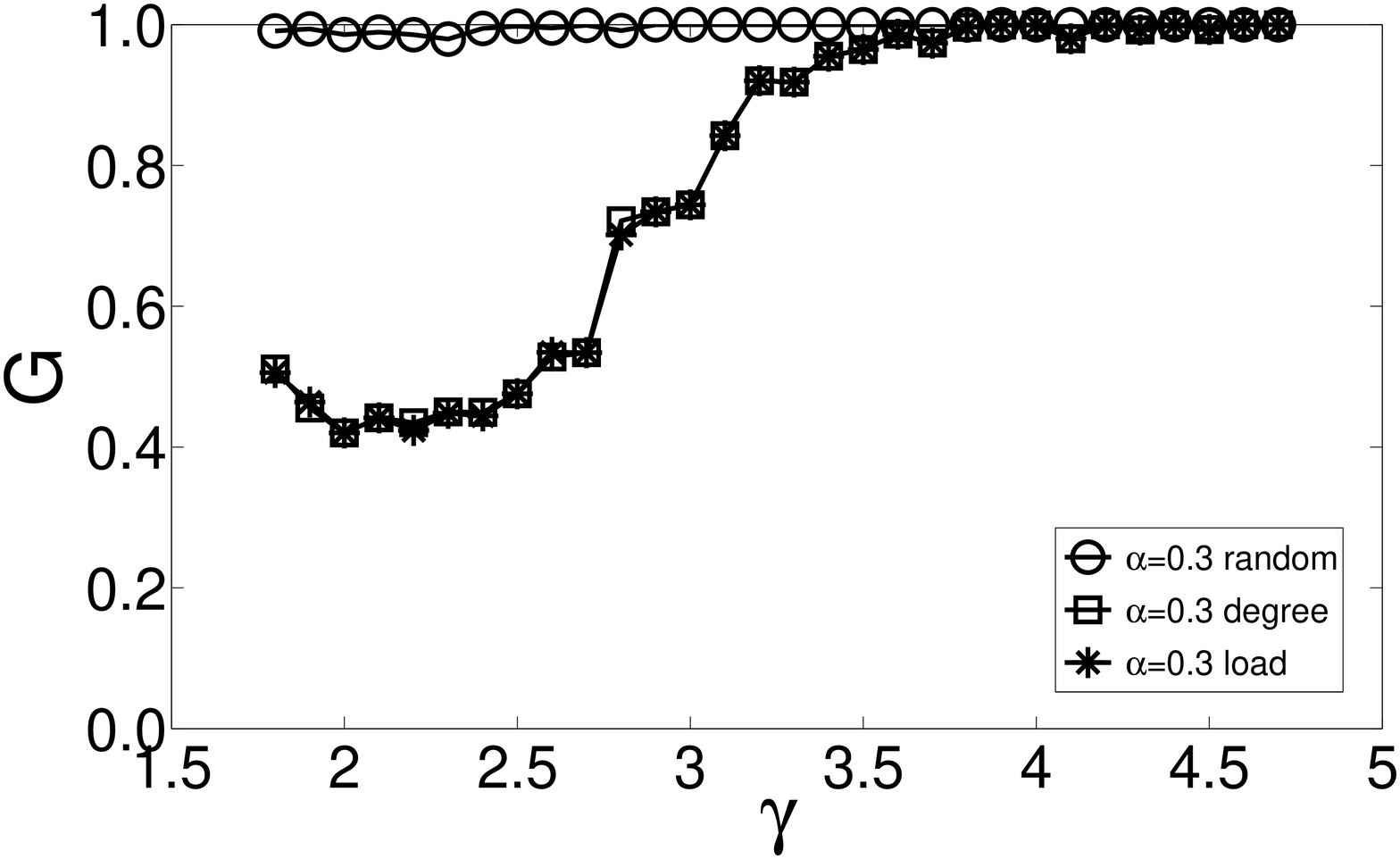} \\
(a)\ \ \ \ \ \ \ \ \ \ \ \ \ \ \ \ \ \ \ \ \ \ \ \ \ \ \ \ \ \ \ \ (b)\\
\includegraphics[scale=0.13]{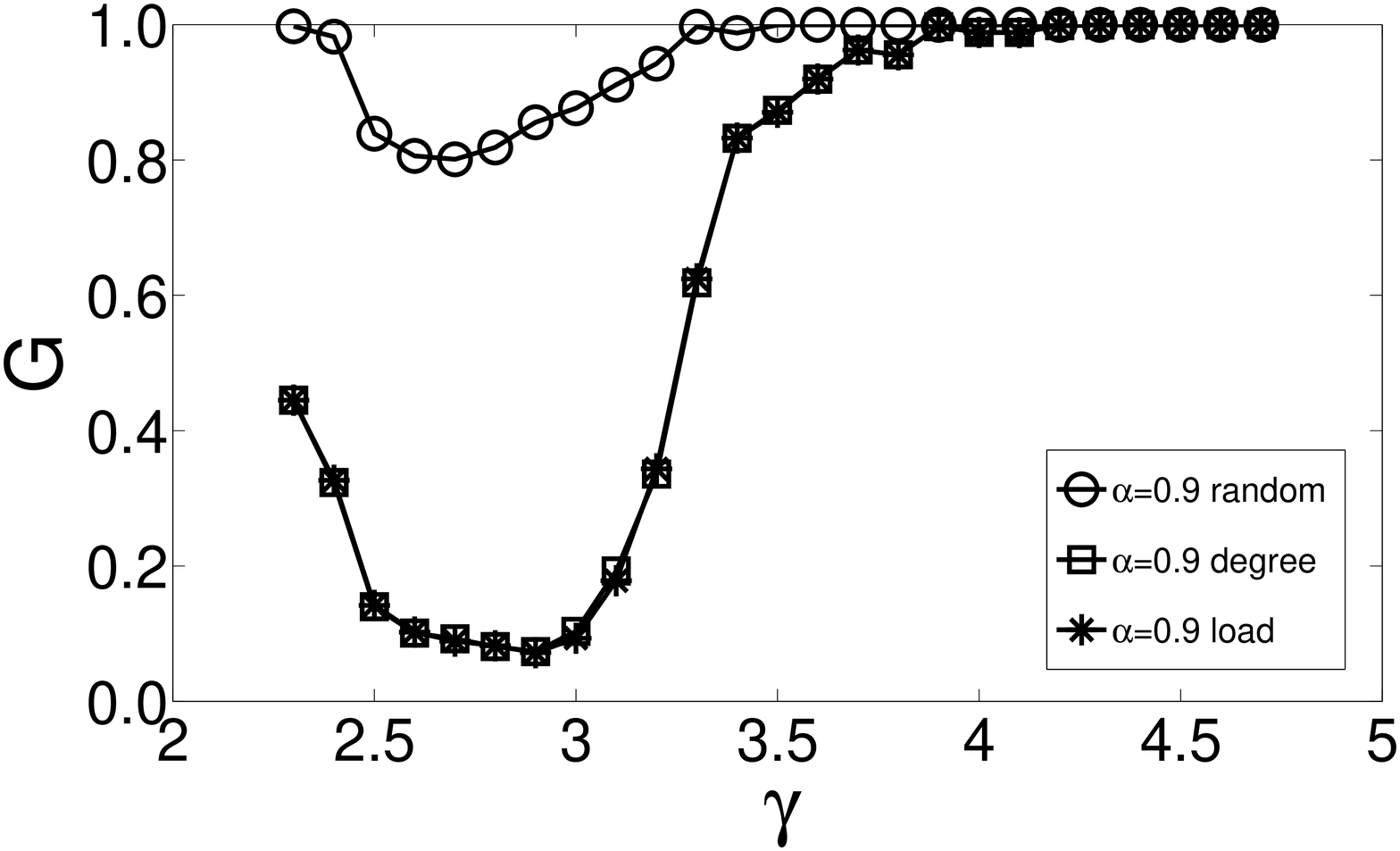}\includegraphics[scale=0.13]{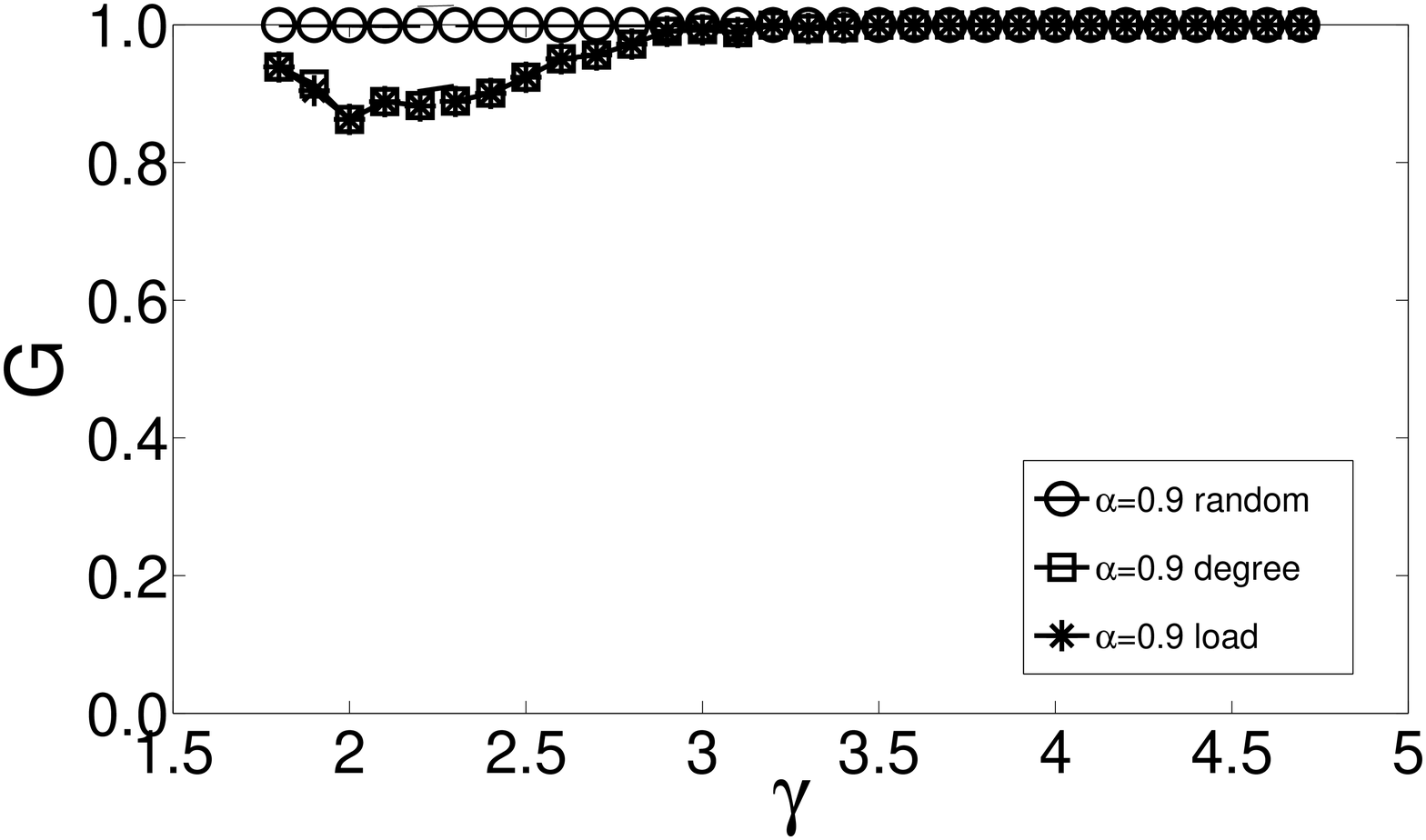} \\
(c)\ \ \ \ \ \ \ \ \ \ \ \ \ \ \ \ \ \ \ \ \ \ \ \ \ \ \ \ \ \ \ \ (d) \\
\includegraphics[scale=0.13]{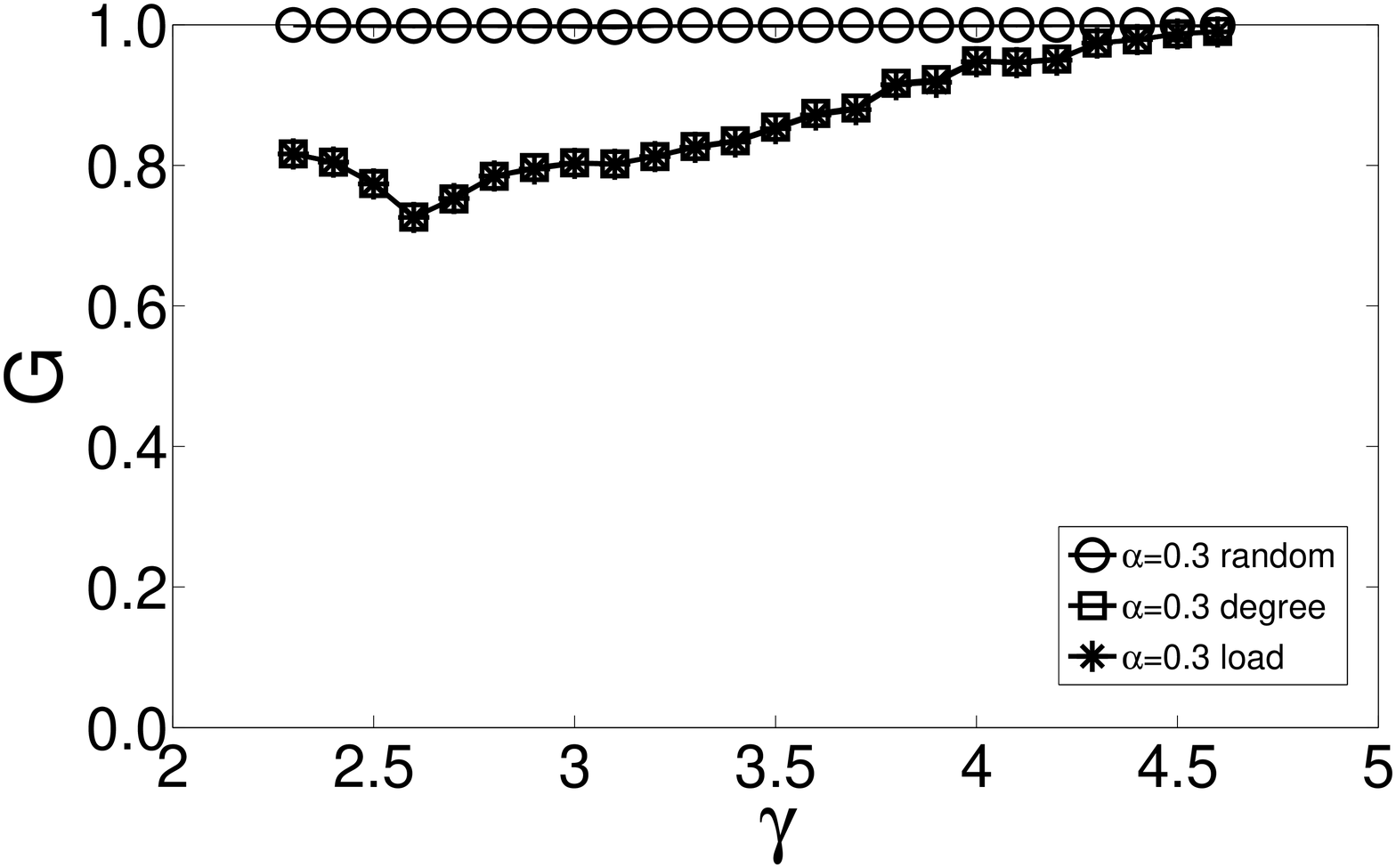}\includegraphics[scale=0.13]{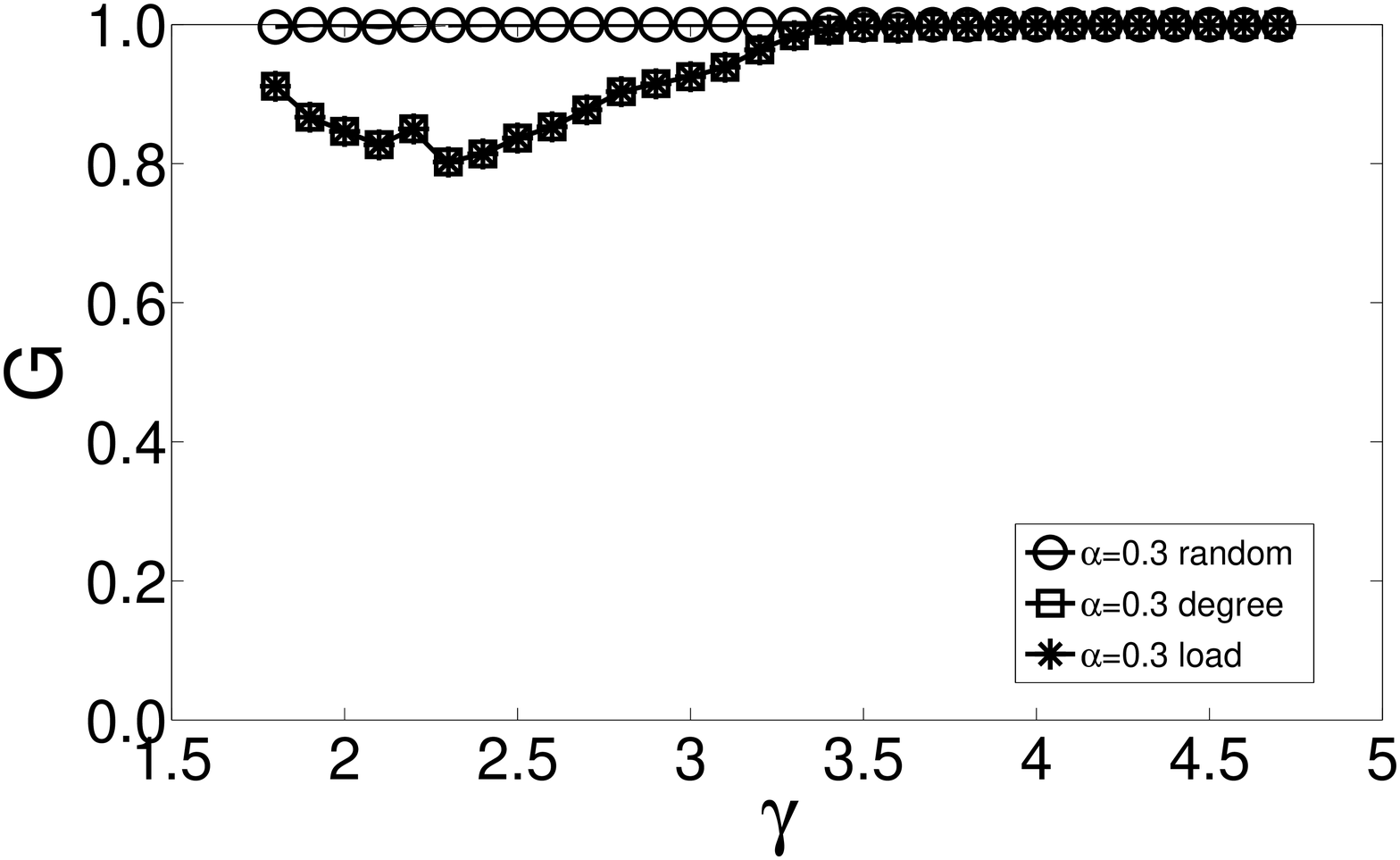} \\
(e)\ \ \ \ \ \ \ \ \ \ \ \ \ \ \ \ \ \ \ \ \ \ \ \ \ \ \ \ \ \ \ \ (f)
\caption{(a) The relative size of giant component $G$ of interdependent SF networks as a function of the scaling exponents $\gamma$, when $N_{A}=N_{B}=5000$, $<k>\approx4$ and $q=1$ for the tolerance parameter $\alpha=0.3$ under different attack strategies. (b) The relative size of giant component $G$ of interdependent SF networks as a function of the scaling exponents $\gamma$, when $N_{A}=N_{B}=5000$, $<k>\approx14$ and $q=1$ for the tolerance parameter $\alpha=0.3$ under different attack strategies. For small degree ($<k>\approx4$), there exits a 'valley' of $\gamma$, while this valley becomes smaller for large degree ($<k>\approx14$). Similar pattern has also found for $\alpha=0.9$ and single networks $q=0$. (c) All the parameters other than $\alpha=0.9$ are the same as (a). (d) All the parameters other than $\alpha=0.9$ are the same as (b). (e) All the parameters other than $q=0$ are the same as (a). (f) All the parameters other than $q=0$ are the same as (b).}\label{4-0.3}
\end{figure}

Network heterogeneity is found to be one of the main causes for cascading failures \cite{16}. The scaling exponent $\gamma$ of scale-free network with degree distribution $P(k)\sim k^{-\gamma}$ can characterize the heterogeneity in the degree distribution. We study the interdependent scale-free network resilience against attacks with different $\gamma$ values in Fig.~\ref{4-0.3}. It is shown in Fig.~\ref{4-0.3} (a) that for $2.5<\gamma<3.7$ of interdependent scale-free network with average degree $<k>\approx4$, intentional attack on one node with highest degree (load) will induce the full collapse of the whole interdependent networks. Even random removal will cause network damage over $60\%$. This demonstrates the extreme network vulnerability as a result of interaction between two failure mechanisms: overloads and interdependency losses. For $<k>\approx4$ we found that there exists a 'valley' of $\gamma$ between $2.5$ and $3.7$ where interdependent network gets minimal resilience. Under random removal, network also reached its minimal resilience around $\gamma=3$. For larger tolerance $\alpha=0.9$ in Fig.~\ref{4-0.3} (c), similar pattern has also been found with a smaller range of the valley for minimal network resilience, where $G$ is comparatively large due to larger system tolerance $\alpha$.

In order to check if this pattern depends on average degree, we increase the average degree largely ($<k>\approx14$). With the finding of a narrow 'valley' of $\gamma$, it is suggested in Fig.\ref{4-0.3} (b) and (d) that interdependent SF networks can preserve larger fraction after attacks and become more robust for large average degree. Effect of large degree can be interpreted as increase of redundancy of paths between pairs, which release the pressure of overloads. Comparing the results shown in Fig.~\ref{4-0.3} (a)-(d), the resilience of interdependent network depends significantly on $\gamma$ values and average degree $<k>$ of the degree distribution $P(k)\sim k^{-\gamma}$.

Compared with single network, the study will help us to identify the effect of interaction between two failure mechanisms in interdependent networks. For single networks (coupling strength $q=0$ in our model) in Fig.~\ref{4-0.3} (e) and (f), the intentional attack can make much less damage to single network compared the interdependent networks in Fig.~\ref{4-0.3} (a) and (b). The difference is also very significant for random removal, which causes almost no damage to single network. Without the interaction of two failure mechanisms, the single network becomes more robust than the interdependent networks.

\begin{figure}
\center
\includegraphics[width=8cm]{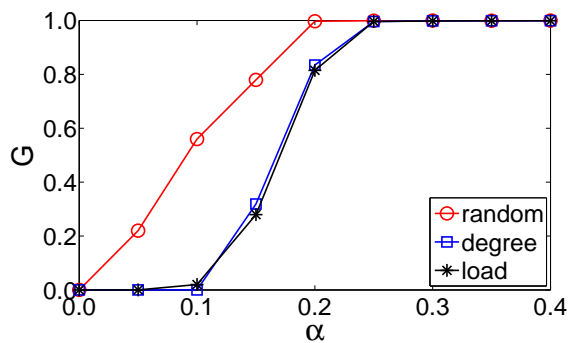}
\caption{The relative size of giant component $G$ of interdependent ER networks as a function of the tolerance parameter $\alpha$, when $N_{A}=N_{B}=5000$, $<k>=4$ and $q$=1 under different attack strategies.}\label{ER}
\end{figure}

For the homogeneous (ER) interdependent networks, Figure.~\ref{ER} shows that when $\alpha=0$, removing one node will break down the whole system, no matter which kind of strategy the removal is based on. At $\alpha=0.1$, it is shown that targeted attacks could damage the largest connected component by more than $90\%$. As the tolerance $\alpha$ increases further (above $0.25$), the removal strategies will hardly trigger a significant breakdown in the networks. It is suggested that the resilience of interdependent ER networks changes more abruptly as $\alpha$ increases compared to interdependent SF networks, where narrow betweenness distribution may be the major cause.

\begin{figure}
\center
\includegraphics[width=8cm]{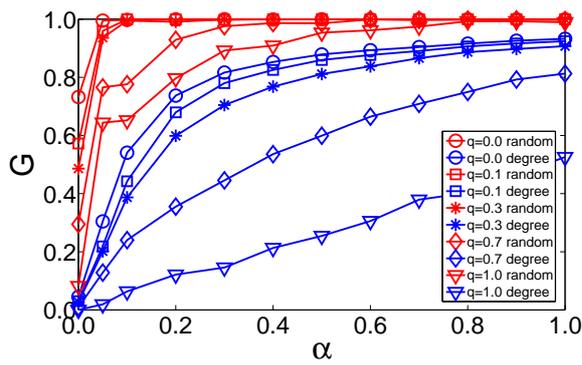}
\caption{The relative size of giant component $G$ of interdependent SF networks as a function of the tolerance parameter $\alpha$, when $N_{A}=N_{B}=5000$, $<k>\approx4$ and $\gamma$=2.3 for four different coupling strength $q$ under random and intentional attack strategies.}\label{sk23}
\end{figure}

After discussing the process of cascading failures in the fully interdependent ER networks and interdependent SF networks with flows, we perform simulations for the case of partial coupling strength $q<1$. In Fig.~\ref{sk23},we present numerical results of the largest component $G$ as a function of $\alpha$ in interdependent scale-free networks under random and intentional attack. It is shown that the network resilience is very sensitive to the interdependency between networks, where interdependency can decrease the network resilience. It is suggested again that interaction between overloads and loss of interdependency changes the network resilience qualitatively.

\section{Conclusion}\label{Conclusion}

Modern complex networks for transporting different flows become more and more dependent on each other. In this paper, we examine the resilience of interdependent networks with flows under random faults and intentional attacks. During the failure processes of interdependent network, there are two possible failure mechanisms: overloads and loss of interdependency. These two mechanisms may amplify each other sometimes, they may slow down each other under some conditions. For example, the nodes removed due to interdependency loss can release the pressure on the network flow and may even reduce the overloads \cite{101}. Due to this complicated interaction between overloads and interdependency loss, interdependent scale-free networks are found to have distinct resilience properties from interdependent networks without flows. For interdependent SF networks with degree distribution $P(k)\sim k^{-\gamma}$, network resilience is not changing monotonically with their scaling exponent $\gamma$, where a valley of minimal network resilience exists in a range of $\gamma$ values. The resilience of interdependent SF networks is found in our simulation much smaller than single SF network or interdependent SF networks without flows. For interdependent ER networks, the resilience changes abruptly possibly due to their narrow betweenness distribution. We believe that further study of interaction between overloads and interdependency losses during the cascading failures processes is an essential step towards fully understanding the resilience of interdependent transportation networks.

\acknowledgments
This work was supported by The National Basic Research Program of China (2012CB725404). Li Daqing would like to add thanks to the support from National Natural Science Foundation of China (Grants No. 61104144). Zhang Peng thanks to the support from National Natural Science Foundation of China (NSFC) under Grants No.11147119 and the Fundamental Research Funds for the Central Universities No.2012RC0707.

\end{document}